\documentclass{ws-procs975x65}

\begin{document}

\title{Cosmic strings with positive $\Lambda$
}

\author{Sourav Bhattacharya $^*$ and Amitabha Lahiri $^\dagger$
}

\address{Department of Theoretical Sciences, S. N. Bose National Centre for Basic Sciences,\\ Block JD, Sector III, Salt Lake, Kolkata 700 098, INDIA\\
$^*$E-mail:sbhatt@bose.res.in,
$^\dagger$E-mail:amitabha@bose.res.in\\
}


\begin{abstract}
We discuss cosmic Nielsen-Olesen strings in space-times
endowed with a
positive cosmological constant. For the cylindrically symmetric, 
 static free cosmic string, we discuss
the contribution of the cosmological constant to the angle
deficit, and to the motion of the null/timelike geodesics. For a non-gravitating cosmic string
in a Schwarzschild-de Sitter space-time, we discuss how a thin string can
pierce the two horizons. We also present a metric which describes the exterior
of a self gravitating thin string present in the Schwarzschild-de Sitter space-time. 
\end{abstract}


\bodymatter

\section{Introduction}\label{s1}
In this talk we discuss cosmic Nielsen-Olesen strings in space-times  endowed with a positive 
cosmological constant $\Lambda$. By Nielsen-Olesen string we mean a vortex line for the Abelian Higgs model~\cite{Nielsen:1973cs}. For $\Lambda=0$, the static cylindrically symmetric free cosmic string space-time has been studied 
by several authors~\cite{Vilenkin2:2000}. The exterior of such strings are Minkowskian with a deficit in the azimuthal angle
%
%
proportional to the string mass per unit length. It is also known that~\cite{Aryal:1986sz, Achucarro:1995nu} a Schwarzschild black hole can be pierced by a cosmic string and for a self gravitating string the exterior also exhibits a deficit in the azimuthal angle. 

Recent observations suggest that there is a strong possibility that our universe is endowed with a small but positive cosmological constant $\Lambda>0$.~\cite{Riess:1998cb, Perlmutter:1998np}. This implies that any observer would find a cosmological event horizon at a length scale of $\sim {\Lambda}^{-\left(\frac{1}{2}\right)}$. The observed tiny value of $\Lambda$ $\sim$ $ 10^{-52}$ m$^{-2}$ implies the length scale of the cosmological horizon is very large, so one may be tempted to neglect the effect of $\Lambda$ in local physics. However, there are situations when local physics is affected by the global topology of the space-time. This is because when one has a cosmological event horizon one cannot reach the spatial infinity and therefore the boundary conditions which must now be set at the cosmological event horizon may be very different from those of the asymptotically flat space-times. Motivated by this we have investigated cosmic strings in de Sitter spaces. The discussion will mainly be based (except the last part of Sec. \ref{s3}) on an earlier work~\cite{Bhattacharya:2008fu} of us to which we refer our readers for details of technicalities and references.      

\section{Free cosmic string and angle deficit}\label{s2}
We start with the metric describing a cylindrically symmetric and static de Sitter vacuum \cite{Bhattacharya:2008fu, Tian:1986, Linet:1986sr, BezerradeMello:2003ei}
\begin{eqnarray}
ds^2=\cos^{\frac{4}{3}}\frac{\rho \sqrt{3\Lambda}}{2}
\left(-dt^2+dz^2 \right) +\frac{4}{3\Lambda}\sin^2 \frac{\rho
 \sqrt{3\Lambda}}{2}\cos^{-\frac{2}{3}}\frac{\rho
 \sqrt{3\Lambda}}{2}d\phi^2+d\rho^2.
\label{p2}
\end{eqnarray}
 The metric is singular at $\rho=\frac{n\pi}{\sqrt{3\Lambda}}$,
where $n$ are integers. Of these points, those corresponding to
even $n$ are flat, with $n=0$ being the axis. The points
corresponding to odd $n$ are curvature singularities. So, our region of interest is from the axis $\rho=0$ to the first curvature singularity located at $\rho=\frac{\pi}{\sqrt{3\Lambda}}$. Now we wish to construct a cylindrically symmetric Abelian Higgs vortex around the axis 
(from $\rho=0$ to say, $\rho=\rho_{0}$) and far away from the naked curvature singularity (i.e., $\rho_0\ll \frac{\pi}{\sqrt{3\Lambda}}$). Let $X(\rho)$ and $P(\rho)$ be the amplitudes of the complex scalar and the gauge field respectively. The boundary conditions appropriate for the string are $X\to 0,~P\to 1$ as $\rho\to 0$, and $X \to 1,~P \to 0$ sufficiently rapidly outside $\rho_0$. In order to solve Einstein's equations within the core we assume that the core is in the false vacuum entirely, i.e., $X=0,~P=1$ inside the core and $X=1,~P=0$ outside.
Then the metric inside core is the same as (\ref{p2}), except $\Lambda$ is now replaced by $\Lambda^{\prime}=\left(\Lambda+2\pi G \lambda \eta^4\right)$, where $(\lambda,~\eta)$ are the parameters of the Abelian Higgs model. The solution outside is (\ref{p2}) with a constant $\delta^2$ multiplied with $g_{\phi\phi}$. 
%
%
 To evaluate $\delta$ we consider the integral $\frac{1}{2 \pi}\int \int\sqrt{g^{(2)}} d\rho d\phi 
\left(G_{t}\,^{t}+\Lambda\right)$. This integral is nonzero only within the core. We shall not quote here the full expression but merely mention the simplified one for GUT strings : $\delta \approx 1 - 4G\mu\left(1 + \frac34\rho_0^2\Lambda +G\mu\right)$, where $\mu$ is the string mass per unit length. Note that this expression gives the correct limit for $\Lambda=0$ in the leading order.  

For the motion of timelike/null geodesics in this space-time, note that the Killing fields define the conserved quantities associated with the geodesics. Using these one can approximately integrate the expression for $\frac{d\phi}{d \rho}$, giving the attractive effects due to the mass within the core as well as the repulsive effects due to positive $\Lambda$~\cite{Bhattacharya:2008fu}.

\section{Cosmic string and the Schwarzschild-de Sitter black hole}\label{s3}
First we consider a non gravitating and cylindrically symmetric Abelian Higgs 
distribution in Schwarzschild-de Sitter background. We choose the string boundary condition as before. 
We assume that the string core width is small compared to the black hole and the cosmological horizon (thin string), and also that the black hole horizon is small compared to the cosmological horizon due to the small value of $\Lambda$. For winding number $N$, the core width is approximated by its flat space value $\sim\frac{\sqrt{N}}{\eta \sqrt{\lambda}}$. If we now expand the equations of motion for the Higgs and gauge fields in the Schwarzschild-de Sitter background within the two horizons, at the leading order these reduce to those in the Minkowskian space-time, solutions of which are known to exist~\cite{Nielsen:1973cs}. To analyze these equations around the horizon we construct two Kruskal like patches around the two horizons. With these patches and the approximations made above one again obtains Nielsen-Olesen equations in the leading order~\cite{Bhattacharya:2008fu}.      

Finally we outline how we may solve for the exterior of a self gravitating Abelian Higgs string present in the Schwarzschild-de Sitter space-time. The procedure of iteration, using Weyl coordinates~\cite{Achucarro:1995nu} may not be applicable in the presence of a positive $\Lambda$. But this problem may be solved without constructing any particular class of axisymmetric coordinates. The basic idea is the following. For the Schwarzschild-de Sitter space in spherical coordinates $(t,~r,~\theta,~\phi)$, we define a transverse radial coordinate $R=r\sin\theta$ defining the string core. When the string is very thin compared to the black hole we have $R\ll r$ inside the core. Then inside the core we may take new coordinates $(t,~r,~R,~\phi)$. Note that basically we are replacing the polar angle $\theta$ by $R$ inside the core. An ansatz can then shown to be made inside the core for thin string
\begin{eqnarray}
ds^2= -A(r,~R)dt^2+B(r,~R)dr^2+dR^2+C(r,~R)d\phi^2.
\label{ans}
\end{eqnarray}
Next we use the Killing identity for the azimuthal Killing field $\phi^a$ : $\nabla_a\nabla^a\phi_b=-R_{ab}\phi^a$. We contract both sides of this identity by $\phi^b$ and  use Einstein's equations $R_{ab}=8\pi G  \left(T_{ab}-\frac{1}{2}T g_{ab}\right)+\Lambda g_{ab}$. When we assume that the string core is essentially in the false vacuum of the complex scalar, we can show that $\left(T_{ab}-\frac{1}{2}T g_{ab}\right)\phi^a\phi^b\approx8 \pi G T_{t}{}^{t}g_{\phi\phi}$.  We may then obtain in the leading order the exterior (with string radius $R_0$)
\begin{eqnarray}
ds^2= -\left(1-\frac{2M}{r}-\frac{\Lambda r^2}{3}\right)dt^2+
\left(1-\frac{2M}{r}-\frac{\Lambda
r^2}{3}\right)^{-1}dr^2+r^2d\theta^2+\left(1-4G\mu-\frac{\Lambda {R_0}^2}{2}\right)^2 r^2\sin^2\theta d\phi^2.\nonumber\\
\label{p4}
\end{eqnarray}
 This generalizes the result of~\cite{S.Deser:1984} for $3-$d de Sitter space without black hole. The limit $R_0 \to 0$ gives the result for a $\delta$ function string.


\end{document}